\documentclass[12pt,preprint]{aastex}
\newcommand{\hii}{\ion{H}{2}}
\begin{document}


\title{$ISO$ SWS OBSERVATIONS OF H~II REGIONS IN NGC~6822 AND I ZW 36:
SULFUR ABUNDANCES AND TEMPERATURE FLUCTUATIONS\altaffilmark{1}}

\author{Joshua G. Nollenberg, Evan. D. Skillman}
\affil{Astronomy Department, School of Physics and Astronomy,
	University of Minnesota, 116 Church St. S.E.,
	Minneapolis, MN 55455}
\email{manawa@astro.umn.edu, skillman@astro.umn.edu}
\author{Donald R. Garnett}
\affil{Steward Observatory, University of Arizona, 933 N. Cherry
	Ave., Tucson, AZ 85721}
\email{dgarnett@as.arizona.edu}
\and
\author{Harriet L. Dinerstein}
\affil{Department of Astronomy, University of Texas at Austin,
	Austin, TX 78712}
\email{harriet@astro.as.utexas.edu}

\altaffiltext{1}{Based on observations with ISO, an ESA project with
instruments funded by ESA member states (especially the PI countries:
France, Germany, the Netherlands, and the United Kingdom) and with the
participation of ISAS and NASA.}

\begin{abstract}
We report $ISO$ SWS infrared spectroscopy of the H~II 
region Hubble~V in NGC 6822 and the blue compact dwarf galaxy I~Zw~36. 
Observations of Br$\alpha$, [S~III] at 18.7 and 
33.5$\mu$m, and [S~IV] at 10.5$\mu$m are used to determine ionic sulfur 
abundances in these H~II regions. There is relatively good agreement between 
our observations and predictions of S$^{+3}$ abundances based on photoionization 
calculations, although there is an offset in the sense that the 
models overpredict the S$^{+3}$ abundances. We emphasize a need for more 
observations of this type in order to place nebular sulfur abundance 
determinations on firmer ground.
The S/O ratios derived using the $ISO$ observations in 
combination with optical data are consistent with values of S/O,
derived from optical measurements of other metal-poor galaxies.

We present a new formalism for the simultaneous determination of 
the temperature, temperature fluctuations, and abundances in a nebula, 
given a mix of optical and infrared observed line ratios. 
The uncertainties in our ISO measurements and the 
lack of observations of [S~III] $\lambda 9532$ or $\lambda 9069$
do not allow an accurate 
determination of the amplitude of temperature fluctuations for
Hubble~V and I~Zw~36.
Finally, using synthetic data, we illustrate the diagnostic
power and limitations of our new method.

\end{abstract}


\section{Introduction}
\label{intro}

Because of their low metallicities 
\citep{PE81, SKH89, IT99},
dwarf irregular and blue compact galaxies can provide valuable 
information for a wide variety of astrophysical problems.  
By comparing the low metal abundances found in dwarf galaxies to 
abundances found in luminous spirals, one can infer variations in 
star formation histories and chemical evolution. It is possible to 
use measurements of abundances in H~II regions in dwarf irregular 
galaxies to establish limits on yields from stellar and big bang 
nucleosynthesis \citep{Pe92}.  It is also possible to 
characterize the ionizing radiation from OB stars from measurements 
of emission lines in H~II region spectra without resolving individual 
stellar spectra \citep{VP88}. 

Heavy elements such as C, N, O, Ne, S, and Ar are typically observed in 
H~II regions.  In order for accurate abundances to be determined, it is
necessary to observe all of the ionization stages present in an H~II
region, or to have a reliable method for inferring the contribution
of unobserved ions.  Of the aforementioned elements, oxygen is the only
one for which all important ionization stages can be easily observed 
at optical wavelengths. In the case of sulfur, the primary optical lines 
are [S~II] $\lambda\lambda$6717,6731 and [S~III] $\lambda$6312. However, 
the [S~III] $\lambda$6312 line is an extremely weak and temperature
sensitive line, making it difficult to measure in many H~II regions. 
Because a significant fraction of the sulfur in an H~II region is in 
the ionization state S$^{+2}$ \citep{G89}, accurate determination of 
S/H = N(S)/N(H) can be difficult based on optical measurements alone. 
The [S~III] $\lambda\lambda$9069,9532 lines in the near-infrared, which 
are intrinsically much stronger, often require a correction for 
atmospheric water absorption. In high-ionization nebulae S$^{+3}$, 
which emits only in the 10.5$\mu$m line, can also become an important 
constituent.  

Furthermore, the depth of particular ionization zones, temperature 
fluctuations ($t^2$), and other variations throughout a nebula can 
cause optical/UV forbidden line diagnostics to yield temperatures 
that are larger than ion-weighted average values in photoionization 
models, and they can weight emissivities toward values found in higher 
temperature regions of the nebulae \citep{P67, G92, MTP98, Ee99}. 
These potential problems motivate the use of temperature insensitive 
mid- and far-infrared forbidden fine structure transitions in order 
to include optically unobserved ions, accurately determine nebular 
abundances, and to determine the amplitude and scale of temperature 
fluctuations inside H~II regions (e.g., Dinerstein, Lester \& Werner 
1985). The infrared portion 
of the spectrum contains strong emission lines such as [S~III] 18.7$\mu$m
and [S~IV] 10.5$\mu$m. For faint extragalactic H II regions,
the high background flux from earth's atmosphere precludes ground-based
observations of many middle- and far-infrared emission lines at the present
time. However, the low background and high sensitivity of the ISO observatory
allowed observations to be made of many faint extragalactic sources.
 
In this paper, we present mid-infrared $ISO$ spectra of the [S~III] 
18.7 $\mu$m and 33.5 $\mu$m lines, plus the [S~IV] 10.5 $\mu$m line, 
from the H~II region Hubble~V in NGC~6822 (hereafter referred to as 
Hubble~V), and the blue compact galaxy I~Zw~36 (MRK~209; UGCA~281). 
Our goal is to compare our infrared observations with published optical 
observations in order to test whether the theoretically predicted
ionization correction factors are correct and to determine whether
temperature fluctuations are large enough to be detectable in this 
manner.

\section{Observations}
\label{obs}

\subsection{$ISO$ Observations}
\label{iso}

Observations of Hubble~V and I~Zw~36 were obtained using the Short 
Wavelength Spectrometer (SWS) \citep{dG96} on the 60 cm Infrared 
Space Observatory ($ISO$) \citep{K96}. Details of these observations are given 
in Table 1. The Astronomical Observation Template (AOT) AOT02 was 
used to measure individual lines in a narrow bandpass with a width 
of $\Delta\lambda$/$\lambda$ $\approx$ 0.01 \citep{L97}. Although 
line profiles were oversampled with the SWS, the instrumental
profile FWHM of about 150 km s$^{-1}$ is much larger than the 
intrinsic widths of emission lines from H~II regions.
The Br$\alpha$ and [S~IV] 10.5$\mu$m lines were
observed through a 14$\arcsec$$\times$20$\arcsec$ aperture, 
while the [S~III] 18.7$\mu$m line was measured through
a 14$\arcsec$$\times$27$\arcsec$ aperture. We also obtained
measurements of [S~III] 33.5$\mu$m through a 20$\arcsec$$\times$33$\arcsec$ 
aperture, but these observations had poor signal/noise and were not used.

Standard $ISO$ reduction techniques were employed to reduce the data,
using the latest photometric calibrations and procedures (SWS ia3) available at
the time of the data reduction \citep{L97}. The following paragraphs
illustrate some of the difficulties and challenges involved in the 
reduction of the $ISO$ data. 

Each $ISO$ observation begins with a photometric check, which is a
detector scan of an internal calibration source.  This is followed by a
series of dark current scans and integrations on the source. 
Memory effects, fringing, glitches and floating dark current levels
seriously degraded the quality of the data that were obtained with the 
SWS. Each of these problems were evident upon preliminary review of the 
data.  Therefore, it was necessary to work with staff scientists and 
programmers at IPAC to correct the data by using the Interactive Analysis 
packages that have been developed.

The memory effects were primarily due to the measurement of the internal
calibration sources at the beginning of each observation.  The intensity
of the calibrators was high enough to cause the detectors to produce
artificially high readings while performing subsequent dark scans and
observations.  This latent signal that persisted
after illumination by bright sources was primarily
evident in Detector Bands 1 and 2.  During the Interactive Analysis,
memory effects were corrected first, using the {\it antimem} IDL routine
that was designed specifically for this purpose.

The second effect that we corrected for was the variation in dark current
readings.  At regular intervals while observing an object, the detector
would perform a dark current scan.  Because of time-dependent and 
nonlinear detector response effects (e.g., memory effects, changing
noise levels, and ``glitches''), dark current subtractions by the 
automated processing pipeline were often inaccurate, especially for
very faint continuum sources. The dark 
scans were corrected by hand using the {\it dark\_inter} algorithm.
First, each individual dark scan was sigma clipped about the median of
the scan using a threshold of $3\sigma$.  
In severe cases, entire dark scans were thrown out (especially 
dark scans that suffered from memory effects because they immediately 
followed a photometric check). 
Individual scans were also deglitched, as will be discussed later. 
While making the dark current corrections, it was necessary to determine
whether the dark current level, which appeared to jump at random time
intervals, actually correlated with the subsequent object scan.  In cases
where there was bimodality, in which the voltage readout fluctuated 
between two levels from one readout point to the next on very short 
timescales, only those points which corresponded to the dark 
level of the object scans were used.  Use of these corrected dark levels 
resulted in a marked improvement of the quality of dark subtraction.

Variations in sensitivity were also a problem in the $ISO$ detectors 
\citep{L97}. 
Generally, these variations were time-dependent and nonlinear.  Variations
from orbit to orbit were often caused by passage through the Van Allen
Radiation Belts, cosmic ray strikes and observation of bright objects.  
However, small variations could be corrected using the {\it spd-rl} 
routine in the Interactive Analysis package. 

Once the above operations using the Interactive
Analysis package were completed, the data were processed further using the 
$ISO$ Spectral Analysis Package (ISAP).  
First, glitches flagged by the initial pipeline processing were removed 
using the algorithm provided in ISAP. 
Generally, about 30\% of the total number
of data points were rejected including some entire scans. 

Another concern is whether any flux was inadvertently lost or gained due
to $ISO$'s nominal $\approx$ 1$^{\prime\prime}$ pointing error. The 
SWS apertures cover an area larger than I~Zw~36, which has an angular 
extent of roughly $10^{\prime\prime}$ x $11^{\prime\prime}$ as well as 
Hubble~V, with a FWHM of $5.5^{\prime\prime}$  \citep{CHK95}. A thorough 
review of the current readouts from each observation showed no systematic
variation with time, only occasional spikes due to cosmic ray hits and 
noise. Variations in the average current level would have indicated that 
the objects were falling in and out of the aperture during the scans.  
This implies that pointing error or jitter did not cause the objects to 
fall out of the aperture. The observed $ISO$ line fluxes for I~Zw~36 and 
NGC 6822 are given in Table 2, and plots of the spectra can be found in 
Figure 1.  

Estimates of the magnitudes of flux calibration uncertainties were given
in Leech (1997) for several of these problems.  Uncertainties due to the
memory effects were reported to be on the order of $6 - 15\% $ in Band 2 
(Br$\alpha$) and $8 - 30\% $ in Band 4 ([S~III] $18.7\mu m$ and
[S~IV] $10.5\mu m$).  Therefore there is a potential uncertainty of up to
nearly $50\% $ in the relative flux calibrations between bands.  
Furthermore, the Spectral Energy Distributions of standard objects used
in flux calibrations are known only to between $4 - 10\% $.
These systematic uncertainties are not included in the line flux errors 
cited in Table 2.

\subsection{Supporting Observations from the Literature}
\label{ukirt}

There exist several sources of published optical data for Hubble~V
\citep{Le79,PES80,ST89,M96} and I~Zw~36
\citep{VT83, ITL97} which allow a comparison of the abundances derived
from optical and infrared [S~III] transitions.  Observed values of
line ratios relevant to this paper
are given in Table 3.  A comparison of the abundances derived from
these sources with the abundances derived from our $ISO$ observations
can be found in \S 3.3.  Of the Hubble~V observations, only
those of Lequeux et al. (1979) include a measurement of the [S~III] 
$\lambda 6312$ line through a large aperture. We therefore adopt these 
optical observations for comparison with our infrared measurements.
Note that the reddening corrections for the infrared emission line ratios 
used in this paper are negligible.

\section{Data Analysis}

\subsection{Electron Temperatures and Densities}

Prior to determining the ionic abundances from the data, electron
temperatures were computed using a combination of existing published 
optical diagnostic data and updated atomic data from \cite{PP95}.  
In the case of Hubble~V, four sources had
[O~III] data: Pagel \& Edmunds (1981), Lequeux et al. (1979), Skillman, 
Terlevich, \& Melnick (1989), and Miller (1996).  From 
these sources, the diagnostic ratio \begin{equation} R(O~III) =
\frac{I(\lambda4959)
+ I(\lambda5007)}{I(\lambda4363)} \end{equation} was determined.
These line ratios and the derived electron temperatures are given 
in Table 4.  Adopting the Lequeux et al. (1979) data, the electron 
temperature for Hubble~V was determined to be $T_e$ = 11,200$\pm$1,100 K.  
This value is consistent, within stated errors, with the other observations
and was adopted for the calculation of emissivities in the 
abundance calculations.  Viallefond \& Thuan (1983) and Izotov, Thuan \&
Lipovetsky (1997) have reported optical spectroscopy for I~Zw~36.  
Using the emission line fluxes from Viallefond \& Thuan (1983), we obtain 
an electron temperature of 14,600$\pm$500 K, a result which corresponds 
closely to that given by
Viallefond \& Thuan (1983): $T_e$ = 14,500$\pm$1,300 K.  However, using
fluxes from Izotov, Thuan \& Lipovetsky (1997) with smaller relative
observational errors, we obtained a value of $T_e$ = 16,180$\pm$70 K,
which we adopt.

Photoionization models indicate that the electron temperature is not
uniform across an H~II region, but can be different for low-ionization
and high-ionization zones, depending on metallicity (Garnett 1992).
We use the formulation given in Garnett (1992) to estimate electron 
temperatures for the O$^+$ and S$^{+2}$ zones, based on 
the derived [O~III] temperature. We derive T(O$^+$) = 10,800$\pm$1,200 
K for Hubble~V and T(O$^+$) = 14,330$\pm$50 K for I~Zw~36, while 
T(S$^{+2}$) = 11,000$\pm$1,200 K and 15,130$\pm$50 K for Hubble~V and 
I~Zw~36, respectively.  Since the temperatures in the low ionization
zones are not observed directly, but rather are derived from photoionization
models, we assume a lower limit of 500 K on 
the uncertainty when deriving abundances.

Published spectroscopic results for these two regions give low 
electron densities, $n_e$ $<$ 200 cm$^{-3}$ based on [S~II] line
ratios. These densities are too small to cause significant collisional 
de-excitation of the [S~III] and [S~IV] lines. Because of the poor 
signal/noise in the 33.5$\mu$m observations, these infrared [S~III]
line measurements do not provide meaningful constraints on $n_e$. 

\subsection{Abundance Calculations}

Using the values of electron temperature derived in the previous
section, we computed ionic abundances using a 5-level atom for 
O$^+$, O$^{+2}$, S$^+$, and S$^{+2}$ from the published optical 
emission line data for Hubble~V and I~Zw~36. Collision strengths for 
[O~II], [O~III], and [S~II] transitions were taken from the compilation
of Pradhan \& Peng (1995). Collision strengths for [S~III] were taken
from the 27-state R-matrix calculation of Tayal \& Gupta (1999), while 
for [S~IV] we took the results of the 24-state calculation of Tayal 
(2000). Our computed abundances are 
listed in Table 5.  We note that the new values of the effective 
collision strengths for the [S~III] infrared fine-structure transitions 
represent an increase of $\approx$ 30\% over those of \cite{GMZ95}.  

S$^{+3}$ can represent a substantial fraction of the total sulfur 
abundance in many H~II regions (e.g., Lester, Dinerstein, \& Rank
1979; Pipher et al. 1984). \cite{G89} (see also Garnett et al. 1997) 
showed that there is a sharp decrease in the observed ratio 
(S$^{+} +$S$^{+2}$)/O for O$^+$/O $<$ 0.25, indicating an increasing 
contribution from S$^{+3}$ in high ionization H~II regions 
(where ionization is parameterized by O$^{+}$/O). However, the 
contribution of S$^{+3}$ often remains uncertain because [S~IV] 
has no optical transitions; the only line in the ground configuration
is at a wavelength of 10.51$\mu$m. In most cases photoionization 
models are used to estimate the S$^{+3}$ contribution. \cite{D80}
demonstrated that sulfur ionization correction factors based on
coincidences in ionization potentials greatly overpredict the 
actual amount of S$^{+3}$ in high-ionization planetary nebulae.
However, the accuracy of ionization corrections based on 
photoionization models is largely untested for H II regions,
since very few observations exist for
[S III] and [S IV] in comparable beam sizes. 
Therefore, an important aspect of this study is the inclusion of 
[S~IV] in the estimation of the total nebular sulfur abundance.  

Here, we determine S$^{+2}$ and S$^{+3}$ abundances from our $ISO$
observations of the infrared fine structure lines [S~III] 18.7$\mu$m
and [S~IV] 10.5$\mu$m.
We normalized the IR fine-structure lines to our Br$\alpha$ 
measurements also made with the SWS.  Because the extinction
coefficients are low in the IR and even the optical extinction
to these two targets are low, the reddening corrections are
negligible. 
Given the nebular 
physical conditions derived in Section 3.1, we computed the S$^{+2}$
and S$^{+3}$ abundances listed in Table 5.
We used only the 18.7$\mu$m line to derive the S$^{+2}$ abundance,
because of the low signal/noise for our 33.5$\mu$m measurements.

Theoretical models and observational studies have indicated that the 
S/O ratio remains constant with respect to the oxygen abundance, O/H, 
regardless of metallicity \citep{Fr88,TPF89,G89,IT99}. In Figure 2 we 
plot our newly-derived S/O values for Hubble~V and I~Zw~36, along with 
values for other objects obtained from the literature, vs. O/H. Figure 
2 shows that both I~Zw~36 and Hubble~V have S/O values similar to those 
of other metal-poor H~II regions. This result tends to support the
validity of the abundance ratios derived from optical spectroscopy.

\subsection{Photoionization Models}

Current photoionization models are limited in accuracy because of
uncertainties in input parameters such as stellar ionizing flux 
distributions. \cite{VP88} proposed the use of the ratio 
\begin{equation}
\eta = \frac{O^{+}/O^{+2}}{S^{+}/S^{+2}}
\end{equation} 
as an indicator of the ``hardness" of the photoionizing radiation 
field inside a nebula, which can be used to infer the effective 
temperature ($T_*$) of the ionizing cluster. Garnett (1989) showed that 
this is indeed an useful estimator of relative values of 
$T_*$ in nebulae by constructing 
photoionization models using different stellar atmosphere flux 
distributions. However, it may not be possible to derive absolute 
values of $T_*$ from $\eta$, and it begins to lose its sensitivity 
above $T_{eff}$ $\sim$ 45,000K, as shown by Skillman (1989). Values 
of $\eta$ for each nebula were calculated using the O/H ratios as well 
as the (S$^{+}+$S$^{+2}$)/S values from Table 5 and are also listed 
there. Using Figure 6 of Garnett (1989) and Figure 1 of \cite{VP88}
we find that the $\eta$ parameter for Hubble~V 
is consistent with $T_*$ 
$\approx$ 45,000K (using Hummer \& Mihalas 1970 LTE model atmospheres)
while the $\eta$ parameter for I~Zw~36 is consistent with $T_*$
greater than 50,000K.
This is within the range of values of $\eta$ for other giant 
extragalactic \hii\ regions and consistent with excitation by a 
mixture of hot O and B type stars (cf.\ Garnett 1989). 
In the specific case of Hubble~V, Bianchi et al.\ (2001) reproduce the
H-R diagram for the most luminous stars in OB 8, the stellar 
association powering Hubble~V, and it appears that 45,000 K is
a conservative upper limit to the effective temperatures of
the most massive stars in this association. 

Early studies of sulfur abundances in \hii\ regions noted that neglecting 
the contribution of S$^{+3}$ would result in an underestimate of the 
total sulfur abundance (e.g., Stasi\`nska 1978; French 1981).
The photoionization models of Garnett (1989) showed a clear 
relationship between the S$^{+3}$ ionization correction and 
O$^+$/O which is relatively insensitive to stellar effective 
temperature or abundance.  In Figure 3, we 
show O$^+$/O vs. log (S$^{+}$+S$^{+2}$)/S for the two observed nebulae 
and compare them to the models of Stasi\'nska (1990) for two different 
sequences in stellar effective temperature. The model sequences represent 
abundances of 0.1 times solar, similar to the metallicities of our two
objects.  The values for Hubble~V and I~Zw~36 plotted in Figure 3 fall 
at slightly higher values of (S$^{+}$+S$^{+2}$)/S or lower values of 
O$^+$/O than the models. This may indicate that the S$^{+3}$ fraction 
is overestimated in the models, perhaps due to line blanketing not 
accounted for in the stellar atmosphere fluxes.  However, we caution 
against over-interpretation of this offset given the mismatch between 
the apertures for the optical and IR observations and the fact that
the points in Figure 3 are less than 2 $\sigma$ away from the 
model curves. We emphasize that more 
observations of S$^+$, S$^{+2}$, and S$^{+3}$ for the same object 
(preferably with matched apertures) will provide a valuable consistency 
check on the ICF for S$^{+3}$. Corrections for the unobserved S$^{+3}$ 
abundance are now usually carried out based on photoionization modeling 
(e.g., the Thuan, Izotov, \& Lipovetsky 1995 fit to the models of 
Stasi\`nska 1990).  It is still desirable to have observational tests 
of these fits spanning a large range in excitation. 
With the advent of more sensitive IR instruments, it may 
eventually be possible to  characterize the effective temperatures 
of the ionizing stars, and to determine the best nebular 
models to describe a given H~II region.

\section{DIAGNOSTICS AND TEMPERATURE FLUCTUATIONS}
\subsection{Standard Analysis of Temperature Fluctuations}
Because optical collisionally excited line emission is weighted toward 
high temperature regions, abundance measurements based on optical data 
may not provide the true ionic abundance of a species in the H~II region
\citep{P67,G92,M95,SVG97}.  This effect would be most extreme in 
a case where there is a very localized zone of high temperature embedded
in a more extended, lower temperature nebula.  In this case,
a calculation of the S$^{+2}$ abundance from [S~III] $\lambda$6312 
would yield a S$^{+2}$ abundance lower than the true value. However,
this problem can be rectified by observing lines with much smaller 
excitation energies, i.e. infrared fine structure transitions.  Because 
of their lower excitation energy, the volume emissivities for 
such transitions are insensitive to electron temperature, and as a result, 
emission line ratios can be converted to ionic abundances with a smaller 
dependence on temperature variations within the ionized gas (e.g., 
Dinerstein 1986). The measurement of larger abundances of an ion from 
infrared line observations than from optical lines would thus be an 
indication that there may be temperature fluctuations inside a nebula. 
\cite{DLW85} used such an approach to estimate the magnitude of 
temperature fluctuations in several planetary nebulae using a 
combination of infrared and optical [O~III] lines.
Comparison of the S$^{+2}$ abundances in Table 5 show evidence of 
the effect of temperature fluctuations in I~Zw~36 and Hubble~V, 
although this is only significant at the 1-2$\sigma$ level. 
   
Peimbert (1967) first determined the effects of temperature fluctuations 
on the calculation of nebular temperatures themselves through a 
density-weighted ensemble average of the temperature, 
\begin{equation} 
T_o(N_i,N_e) = \frac{\int T(r) N_i(r) N_e(r) d \Omega dl}{\int N_i(r) 
N_e(r) d\Omega dl},  
\end{equation} 
derived from an emission line. This average temperature, used in 
conjunction with an emission temperature, defined by:  
\begin{equation} 
\frac{I_{X_{+p},\lambda_{nm}}}{I_{X^{+p},
\lambda_{n_1m_1}}} = exp\left[ - \frac{\Delta E - \Delta E^*}{kT}\right], 
\end{equation} 
where $\Delta E$ and $\Delta E^{*}$ are the excitation 
energies of the two lines,
can be used to define the root mean square temperature 
fluctuation, 
\begin{equation} 
t^2 = \langle \bigg[ \frac{T(r) - T_o}{T_o} \bigg]^{2} \rangle. 
\end{equation} 
Then, assuming small 
fluctuations, one can perform a Taylor expansion, and relate the 
emission temperature to the average temperature by:
\begin{equation} 
T \approx T_o \left[ 1 + \left( \frac{\Delta E + 
\Delta E^*}{kT_o} - 3 \right) \frac{t^2}{2} \right], 
\end{equation} 
where $ \Delta E \neq \Delta E^* $ and $t^2 \ll 1 $.

An alternative formalism was created by Mathis (1995) in which a
fraction, $C$, of the gas is assumed to be at one temperature, $T_1$,
while the rest of the gas is assumed to be at a second temperature, 
$T_2$.  From this, one can estimate the degree to which
differing amounts of plasma at each temperature can affect the derived
abundances through the optical depth: $d\tau /dT = C \delta (T - T_1) + 
(1 - C) \delta (T - T_2)$, where $d\tau = n_e n(X) ds$.

\subsection{A New Approach to Characterizing Temperature Fluctuations}

We have developed a different approach that can be used to characterize the
average temperature and the root-mean-square temperature fluctuations in
a region.  This method can 
be used with an arbitrary temperature distribution.

Assume the gas in a nebula follows a Gaussian temperature distribution, 
with mean temperature, $T_o$, and a dispersion $\sigma_T$. Let us
also assume that the emission lines that we are observing have 
emission coefficients, at constant density, that can be characterized over
a temperature range of a few times $\sigma_T$ by a quadratic fit,
\begin{equation} 
\epsilon_{X_i, \lambda_i}(T) = a_{X_i, \lambda_i} T^2 + 
b_{X_i, \lambda_i} T + c_{X_i, \lambda_i}, 
\end{equation}  
where $a_{X_i, \lambda_i}$, $b_{X_i, \lambda_i}$, and 
$c_{X_i, \lambda_i}$ are the coefficients of the quadratic fit of the
emission coefficient.  This is justified, because emission coefficients
can usually be accurately approximated by quadratic polynomials over 
temperature ranges of several thousand degrees.
 
Then, the ratio of two emission line intensities ($R$), which can 
be generally written as: 
\begin{equation}  
R = \frac{I_{X_1,\lambda_1}}{I_{X_2,\lambda_2}} = 
\frac{N_{X_1} N_e \epsilon_{X_1, \lambda_1}(T)}{N_{X_2} N_e \epsilon_{X_2, 
\lambda_2}(T)} 
\end{equation} 
can be convolved with the  normalized Gaussian 
temperature distribution so that we have:  
\begin{equation} 
 \frac{I_{X_1, \lambda_1}}
{I_{X_2,\lambda_2}} = \frac{N_{X_1}}{N_{X_2}} \frac{\int \epsilon_{X_1, 
\lambda_1}(T) p(T) dT}{\int \epsilon_{X_2,\lambda_2}(T) p(T) dT}, 
\end{equation}
where 
\begin{equation} 
p(T) = \frac{1}{\sigma_T  \sqrt{2\pi}} exp\left[
\frac{(T - T_o)^2}{2\sigma_T^2} \right]. 
\end{equation}  
Upon integration from $T = -\infty$ to $T = \infty$, the relation becomes 
\begin{equation} 
\frac{I_{X_1,\lambda_1}}{I_{X_2,\lambda_2}} = 
\frac{N_{X_1}}{N_{X_2}} \frac{a_{\lambda_1} \left(\sigma_T^2 + T_o^2 
\right) + b_{\lambda_1} T_o + c_{\lambda_1}}{a_{\lambda_2} \left(
\sigma_T^2 + T_o^2 
\right) + b_{\lambda_2} T_o + c_{\lambda_2}}, 
\end{equation} 
which holds to a good degree of accuracy due to the square-exponential 
behavior of the kernel in the integrand.  However, one should be aware
that in this approximation, the Gaussian kernel, $p(T)$, has some
mean temperature, $T_{o}$, and width, $\sigma_{T}$.  If $T_{o} \gg \sigma_{T}$,
then there are no problems, but if $T_{o} \sim \sigma_{T}$, then a significant 
portion of the kernel may correspond to temperatures $T < 0 K$, which is 
clearly non-physical and the approximation breaks down. When $T_o$ $\ge$ 
3$\sigma_T$, then less than $3\%$ of the kernel will correspond to negative 
temperatures, and the approximation will hold.

Let the line ratio divided by the abundance be given by
\begin{equation} 
\gamma_{12} =
\frac{I_{X_1,\lambda_1}}{I_{X_2,\lambda_2}}\cdot\frac{N_{X_2}}{N_{X_1}}, 
\end{equation} 
then it is possible to solve for the temperature fluctuations,
$\sigma_{T}^{2}$, using the equation  
\begin{equation} 
\sigma_{T}^2 = \frac{(a_{\lambda 1} - \gamma_{12} a_{\lambda 2})T_o^2 + 
(b_{\lambda 1} - \gamma_{12} b_{\lambda 2})T_o + (c_{\lambda 1} - 
\gamma_{12} c_{\lambda 2})} {- (a_{\lambda 1} - \gamma_{12} a_{\lambda 2})}. 
\end{equation}  
We will refer to Equation (13) as the ``diagnostic equation."
Equation (13) yields a locus of points in $T_o$,
$\sigma_{T}^{2}$ space:  a curve for a single, 
fixed value for the observed line ratio, $R$, and, for different values of
$T_o$, different Gaussian temperature distributions of variance 
$\sigma_{T}^{2}(T_{o})$ centered on $T_o$.
So by using 
additional line ratios, one establishes a diagnostic in which it is 
possible to solve explicitly for the ionic temperature and the temperature
fluctuations in the gas (as done for [O~III] by Dinerstein, Lester,
\& Werner 1985).  Furthermore, provided three or more collisionally 
excited lines relative to an H recombination line are available for 
a given ionic species, the abundance of that species can also be 
solved for simultaneously. The calculated temperature variance,
$\sigma_{T}^{2},$ is related to the traditional $t^{2}$ by definition:
\begin{equation} 
\sigma_{T}^{2} = T_{o}^{2} t^{2}, 
\end{equation} 
which holds even if the true distribution of the gas temperature is not 
Gaussian.  

If a case arises in which a Gaussian distribution is not appropriate, 
this method is adaptable so that other (normalized) kernels can be 
applied as well.  For example, 
assuming a normalized rectangle function (a top-hat) temperature 
distribution, with half-width $\delta_T$, given by:
\[ p(T) = \left\{ \begin{array}
	{r@{\quad : \quad}l}
	\frac{1}{2\delta_{T}} & T_{L} < T < T_{U}\\
	0 & T \leq T_{L}, T\geq T_{U}.
	\end{array} \right. \] 
 
\noindent
Then one obtains an equation identical
to (11), but of the form
\begin{equation} 
\frac{I_{X_1,\lambda_1}}{I_{X_2,\lambda_2}} = 
\frac{N_{X_1}}{N_{X_2}} \frac{a_{\lambda_1} \left(\frac{1}{3}
\delta_{T}^{2} + T_o^2 
\right) + b_{\lambda_1} T_o + c_{\lambda_1}}{a_{\lambda_2} \left(
\frac{1}{3}\delta_{T}^{2} + T_o^2 
\right) + b_{\lambda_2} T_o + c_{\lambda_2}}. 
\end{equation}
keeping in mind that $\delta_{T}^{2}$ can be related to $\sigma_T^2$ by:
\begin{equation} 
\sigma_{T}^{2} = \frac{1}{3} \cdot \delta_{T}^{2}. 
\end{equation}  
Other distribution functions that also could be used include the Lorentzian
distribution or the Triangle distribution with a base width of $b_{T}$, 
given by

 \[ p(T) = \left\{ \begin{array}
	{r@{\quad : \quad}l}
	\frac{T - T_o + b_T}{b_{T}^{2}} & T_o - b_T < T < T_o\\
	\frac{-T + T_o + b_T}{b_{T}^{2}} & T_o  < T < T_o + b_T\\
 	0 & otherwise.
	\end{array} \right. \] 

\noindent  The Triangle distribution and a truncated Lorentzian distribution 
which has 
limits placed on the width of the wings at its base also yield equation
(11).    For the Triangle distribution, $\sigma_T^2 = 1/6\ \times\ 
 \delta_T^2 $, while
the truncated Lorentzian possesses a more complicated relationship between
its width, base limits, and $\sigma_T^2$.

\subsection{Behavior of the Diagnostic and Choice of Line Ratios}

Initially, we tested the diagnostic by checking whether it would 
be able to determine the physical properties of a model of a gas 
cloud with an average temperature $T_0$ = 12,000 K, temperature
fluctuations $\sigma_T$ = 2,000 K, and an abundance approximately 
equal to that of Hubble~V.  We used emission coefficients given by 
the STSDAS {\it nebular.ionic} routine \citep{SD95}, each of which 
was fitted using a second-order polynomial. Coefficients for
these fits are listed in Table 6. Then the square root of equation 
(14) was plotted using various line ratios involving [S~III] and
[O~III] forbidden lines and H recombination lines, assuming the 
physical conditions given above, by varying $T_o$
in order to determine $T_o$ and $\sigma_T$
simultaneously.  Figure 4 shows these 
diagnostic lines for the case of the nebula described above.  
The top panel shows diagnostics from combinations of [S~III] and H 
recombination lines while the bottom panel shows diagnostics from 
combinations of [O~III] lines.  Diagnostics involving different
combinations of emission lines have different sensitivities to
temperature and temperature fluctuations.  In order to simultaneously
determine the temperature and the temperature fluctuations one would 
ideally use two combinations of emission lines that yield diagnostics
that intersect at nearly right angles. For example, this is roughly 
the case in the
top panel of Figure 4 between the [S~III] $\lambda 6312$/$18.7\mu m$ 
and the  $\lambda 9532$/$18.7\mu m$ diagnostics.

The models on which Figure 4 is based show an additional feature of the 
diagnostic.
The diagnostics involving lines originating solely from one ion
intersect at a point that is independent of ionic abundance.  However,
when using diagnostics that involve ratios of ionic to
H recombination lines, this is no longer the case.  Therefore, it is
possible to simultaneously determine the ionic abundance as well as
the temperature and temperature fluctuations, provided that strengths 
of enough emission lines are known so that the three unknowns can 
be calculated, and that the ions in question reside in regions of
the same $T_0$ and $\sigma_T$.    

Because the diagnostics are comprised of ratios of emission lines
with different temperature sensitivities (the $\lambda 9532$/$18.7\mu m$ 
diagnostic, for example), the sensitivity of the diagnostic will vary 
with temperature.  Graphically, the diagnostic lines appear to rotate 
as either the temperature or the
temperature fluctuations are varied.  Therefore it may be necessary
to mix and match combinations of emission lines in order to have 
orthogonally-crossing diagnostic lines that provide the most stringent
determination of the nebular parameters.  A general rule of
thumb for most nebular conditions would be to use combinations involving
nebular (transitions between the middle and lowest terms in an ionic
configuration, e.g. [S~III] $\lambda$9532), auroral (transitions between
the highest and middle terms, e.g. [S~III] $\lambda$6312), 
and fine structure features.  

\subsection{Application of the Diagnostic Technique}

An optimal combination of emission lines consisting of the nebular, 
auroral, and fine-structure lines of an ionic species as well as 
hydrogen recombination lines will allow the simultaneous determination 
of the nebular temperature, temperature fluctuations, and the abundance
of the species with respect to hydrogen (see Equation 13). Unfortunately, 
for most H~II regions, published observations do not exist for all of 
these types of transitions (or, as in the present case, they have not 
been observed through matched apertures), so we are unable to use our 
diagnostic technique fully with the present data.
However, in this section we will demonstrate the use of the
diagnostic using two approaches.  First, we will demonstrate the
usefulness of the diagnostic using our 
$ISO$ and ground-based measurements for Hubble~V and I~Zw~36. This 
exercise will not lead to accurate estimates of T, $\sigma_T$, 
and abundances, because of the relatively large uncertainties in
the line ratios and the fact that we did not have an optimal set of 
[S~III] line ratios.
However, it will serve as an illustration of the method on real data.
Next, we will show the ability of the diagnostic to make rough predictions 
of the conditions inside a nebula using synthetic data. 

\subsubsection{Illustration with real data}

For Hubble~V and I~Zw~36 we have measurements of [S~III] 
$\lambda 6312$, $18.7 \mu$m, and H recombination lines. 
Thus, we have essentially only two diagnostic line ratios, which
is not sufficient to determine $T_0$, $\sigma_{T}^{2}$ and the
S$^{+2}$ abundance simultaneously. Another [S~III] line, for example
the nebular [S~III] 9532 \AA\ transition, is needed for a full
solution. 

Nevertheless, we can still provide an instructive example and
set constrains on $\sigma_T$ if we assume that the S$^{+2}$ abundances 
that we calculated from our {\it ISO} [S~III] $18.7 \mu m$ 
observations are correct.  This abundance was based on the measurement
of the electron temperature from the [O~III] lines and an assumption of
no temperature fluctuations.  In principle, this assumption should be
ok because the fine-structure lines have weak sensitivity to temperature
fluctuations.  How bad is this assumption?  From 
equation (4) in Garnett (1992), which is just an update of
Peimbert (1967) equation (15), we can get the ion-weighted mean electron
temperature, T(O$^{+2}$), from the measured T(O~III) and an 
estimate of $t^2$.  For T(O~III) $=$ 11,200 K and a $t^2$ value of
0.04, T(O$^{+2}$) $=$ 10,000 K.  Since S$^{+2}$/H$^+$ is roughly
proportional to T$^{-0.5}$ for the IR lines, for commonly claimed
values of 0.03 to 0.04 for $t^2$, the uncertainty in the S$^{+2}$/H$^+$ 
abundance is less than 10\% (which is smaller than our quoted errors).
Thus, this is probably not a bad assumption to adopt for an
illustrative example. 

The results are shown in Figure 5 for both Hubble~V 
and I~Zw~36, where it should be noted that with the
assumption of the S$^{+2}$/H abundance, we are left with only two
undetermined variables:  $T$ and $\sigma_{T}^{2}$.  This allows us
to work in two dimensions and two diagnostics rather than three.  We
did, however, include a third diagnostic ($\lambda$6312/18.7$\mu m$)
in Figure 5 to ensure that all of the diagnostic lines crossed at the
same point and produced consistent results.

In Figure 5, the diagnostic curves are grouped by line ratio. The 
center line in each group represents the results of equation 13 for 
each observed line ratio; the  parallel curves represent the spread
caused by the $\pm 1\sigma$ observational errors in the line ratios.
The expected value of $\sigma_T$ will lie at the centroid of the
$1\sigma$ error box.  In section 3.2, we derived [S~III] temperatures 
of $11,000\pm1200K$ for Hubble~V and $15,130\pm50K$ for I~Zw~36, 
based on the estimates from [O~III] temperatures. Meanwhile, from
the diagnostics in Figure 5 we obtain T[S~III] $\approx 12,000\pm1,000K$ 
for Hubble~V, and $15,500\pm1,000K$ for I~Zw~36.  
Both of the values of T[S~III] derived from Fig. 5 are in good 
agreement with those derived in Section 3.1. 

The data presented in Figure 5 are consistent with a very large range in
$\sigma_T$ corresponding to values of $t^2$ from 0 up to roughly 0.2, 
which is much larger
than values usually considered.  To provide better constraints on 
$\sigma_T$ we would need to obtain higher quality observations of the
infrared [S~III] lines as well as include observations of [S~III]
$\lambda\lambda$9069,9532.  In the next section we demonstrate the
quality of data necessary for this goal. 

\subsubsection{Illustration with synthetic data}

The approach that was followed in Figure 5 is quite useful for the quick
determination of the nebular conditions, however we have only used a
rudimentary method for determining the errors associated with the
original data.  This problem is exacerbated by the fact that under some
circumstances, the formal simultaneous solution of multiple diagnostics of the
form given in Equation (14) may in fact yield $\sigma_{T}^{2} < 0$. 
This could easily arise from observational uncertainties, but if the
error propagation is carried out properly then one would expect the
solution to be {\it consistent} with positive values for $\sigma_{T}$. 

In order to develop a better understanding of the error analysis, we
developed a Monte Carlo simulation to generate synthetic
observations of line ratios whose values are specified along with
their corresponding observational errors.  The specified line ratios 
and errors are sent through the diagnostic so that each observation 
is plotted as a point on the $\sigma_{T}^{2}$ vs. $T$ plane.  
The intensity of points on the plane is the distribution that arises 
from a large number of observations, smeared by observational errors.  
We have plotted two examples of this in Figure 6.  

The top panel in Figure 6 shows the distribution of points that arises
from [S~III] line ratios corresponding to an isothermal nebula at
$T = 12,000K$, with observational errors of 2\% in optical and near-infrared
lines ratios and 5\% in line ratios with infrared lines.
The peak of the distribution 
lies at the point that would be determined in the absence of 
observational errors. This plot also shows that roughly half
of the realizations in the simulation yield $\sigma_{T}^{2} < 0$.  
However, the contours corresponding to 1 and 2$\sigma$
errors in the determination of the nebular parameters show that the
observations could correspond to small fluctuations.  The 1$\sigma$
error corresponds to roughly $t^2$ $\approx$ 0.04.  Thus, it is
possible to make significant constraints on temperature fluctuations
using combinations of optical, near-infrared, and infrared [S III]
lines.  Similarly, the
bottom panel of Figure 6 shows the distribution of realizations
resulting from inputs corresponding to $T = 12,000K$ and $\sigma_{T} =
2,000K$ ($\sigma_{T}^{2} = 4\cdot10^{6} K^{2}$, which corresponds
to $t^{2} = 0.028$), smeared by the same errors as above. 
It is now apparent that the diagnostic lines resulting from
observations $\pm 1\sigma$ errors do roughly correspond to the $1\sigma$ 
bounds on the determinations of the nebular parameters.  

Two things should be noted here.  First, the low observational errors 
assumed in the calculations are difficult to achieve in practice, and matching
apertures is critical.  Long slit spectra of either spatially unresolved 
sources (e.g., extragalactic) or Galactic HII regions where the 
variations can be traced along the slit are probably best suited for
this type of work.  It is probably best to obtain ratios to H recombination
lines in all cases (i.e., H$\alpha$ in the optical, P8, P9, and P10 in
the near-infrared, and Br$\alpha$ in the infrared) to provide a 
normalization.  Second, it is important to point out that for
different temperature ranges, different ionic species will provide
better constraints.  For example, in the example presented in Figure 6,
a combination of [O~III] lines (with the same magnitude of relative
errors) gives roughly a factor of two stronger constraint ($t^2$ $\le$ 
0.02) on the presence of temperature fluctuations. 

\section{SUMMARY AND CONCLUSIONS}

We have reported new ISO observations of the mid-infrared fine structure
[S~III] and [S~IV] lines.  These lines are of great importance to the 
accurate determination of nebular total sulfur abundances.  This is due
to the strong dependence of higher excitation lines on temperature, along
with the fact that there are no strong optical lines for either of 
these species.  With our observations, we have shown that S$^{+3}$ 
can constitute a large fraction of the total sulfur abundance in 
extragalactic H~II regions. This means that if one is
to determine accurately the total sulfur abundance without model 
dependence as in equation (4), then one must make infrared observations
of the fine structure [S~III] and [S~IV] lines.  Once infrared 
observations are made, it becomes possible to test not only useful 
techniques such as the determination of the radiation ``hardness'' 
\citep{VP88} and the determinations of the total sulfur abundance as
in equation (4) \citep{G89}, but it becomes possible to test the 
accuracy of photoionization models.  Our data suggest that ionization 
corrections for sulfur based on oxygen ion ratios and photoionization
models are valid. 

The presence of temperature fluctuations in nebulae can complicate 
the determination of nebular abundances.  
We have developed a new, generalized diagnostic capable 
of determining the amplitude of these fluctuations by assuming 
a Gaussian temperature distribution in these nebulae, although
this method can be applied to any normalized distribution.  
The uncertainties in our ISO measurements and the
lack of observations of [S~III] $\lambda 9532$ or $\lambda 9069$
did not allow an accurate 
determination of the amplitude of temperature fluctuations for
Hubble~V and I~Zw~36 using our method. 
A significant challenge is presented by combining large aperture
infrared spectra with relatively small aperture optical spectra.
Future long slit spectrographs available with SOFIA and SIRTF
will allow us to overcome this challenge. 
As these more powerful instruments become available, observational 
uncertainties should decrease, allowing a more accurate 
determination of the size of the temperature fluctuations and 
other nebular parameters. 

In the future, one can consider extensions to the present analysis. 
For example, the potential to calculate spatially unresolved density 
fluctuations in a similar mathematical framework remains relatively 
unexplored, although \cite{R89} has considered the biasing of IR 
density indicators by density fluctuations. 
Like temperature
fluctuations, density fluctuations can also affect nebular line ratios,
and they may have significant effects on nebular physical parameters
which must be constrained in order to develop better nebular models.
To this end, application of this diagnostic to published data on a 
large number of H~II regions may be useful in characterizing the 
variances found in H~II regions.  

\acknowledgements

We give special thanks to the Nancy Silbermann, Sergio Molinari, Sarah
Unger and the rest of the IPAC staff for their assistance during and 
after our visit to IPAC during January, 1997. 
We also thank the referee, Manuel Peimbert, for a careful reading of
the manuscript and several suggestions which significantly improved
the paper.
JN, EDS, and DRG 
acknowledge support from JPL contract 961500; DRG also acknowledges 
support from NASA LTSA grant NAG5-7714. HLD's participation was
supported by JPL contract 961543. EDS acknowledges support from NASA
LTSA grant NAG5-9221 and the University of Minnesota.

\clearpage

\begin{deluxetable}{lll}
\tablewidth{0pc}
\tablecaption{Journal of ISO SWS Observations}
\tablehead{
\colhead{}    & \colhead{Hubble~V} & \colhead{I~Zw~36}}
\startdata
Observation date & 17 Apr 1996 & 25 Apr 1996 \\
Observer & DGARNETT & DGARNETT \\
TDT      & 15202208 & 16001210 \\
Integration Time & $588 s$ & $8040 s $ \\
\enddata
\end{deluxetable}

\clearpage

\begin{deluxetable}{lccc}
\tablewidth{0pc}
\tablecaption{Observed SWS Line Fluxes}
\tablehead{
\colhead{Object} &\colhead{Wavelength}& \colhead{Species} &
\colhead{Line Flux}\\
\colhead{} &\colhead{($\mu m$)} &\colhead{} & \colhead{($W/cm^2$)}}
\startdata
Hubble~V&4.05 & Br$\alpha$ & $(1.9\pm0.2)\times10^{-20}$ \\
 &10.51 & [S~IV] & $(7.5\pm0.5)\times10^{-20} $ \\
 &18.71 & [S~III] & $(6.7\pm0.7)\times10^{-20}$ \\
\hline
I~Zw~36&4.05 & Br$\alpha$ & $(1.9\pm0.3)\times10^{-21}$ \\
 &10.51 & [S~IV] & $(1.2\pm0.1)\times10^{-20} $ \\
 &18.71 & [S~III] & $(4.9\pm1.0)\times10^{-21}$ \\
\enddata
\end{deluxetable}

\clearpage

\begin{deluxetable}{llcccccc}
\tablewidth{0pc}
\tablecaption{Published Optical Line Strengths}
\tablehead{
\colhead{$\lambda$ (\AA)} & \colhead{Species} & \multicolumn{4}{c}
{Hubble~V} & \multicolumn{2}{c}{I~Zw~36}\\
\cline{1-8} \\
 & & \colhead{LPRST79} &
\colhead{PES80} & \colhead{STM89} & \colhead{M96} & \colhead{VT83} & \colhead{ITL97}}
\startdata
$3727$ & [O~II] & $1.4$ & $1.5$ &$1.20\pm0.06$& $1.46\pm0.10$& $0.7$ & $0.719\pm0.002$\\
$4363$ & [O~III] & $0.052$ & $0.047$ & $0.057\pm0.012$ & $0.06\pm 0.01$ & $0.12$ &
$0.127\pm0.001$\\
$4861$ & H$\beta$ & $1.00$ & $1.00$ & $1.00$ & $1.00$ & $1.00$ & $1.00$\\
$4959$ & [O~III] & $1.8 $ & $1.6$ & $1.92\pm0.010$ & $1.67\pm0.04 $ & $1.98$ &
$1.960\pm0.003$\\
$5007$ & [O~III] & $5.4 $ & $5.0 $ & $5.93\pm0.030 $ & $4.90\pm0.14$&$6.52$ &
$5.543\pm0.008$\\
$6312$ & [S~III] & $0.014$ & \ldots & \ldots & \ldots & $0.017$ & $0.017\pm0.001$\\
$6717$ & [S~II] & $0.063$ & $0.09$& $0.129\pm0.006$\tablenotemark{a}& \ldots &
$0.042$ & $0.061\pm0.001$\\
$6731$ & [S~II] & $0.045$ & $0.06$&  & \ldots & $0.031$ & $0.045\pm0.001$\\
$c($H$\beta)$ & & $0.8$ & $1.05$ & $0.7$ & $0.3\pm0.1$ & $0.41\pm0.12$ &
$0.00$\\
\enddata
\tablenotetext{a} {sum of blended $\lambda 6716 + \lambda 6730$
(STM89 only)}
\end{deluxetable}

\clearpage

\begin{deluxetable}{lccc}
\tablewidth{0pc}
\tablecaption{Electron Temperatures}
\tablehead{
\colhead{Source}& \colhead{Object}& \colhead{R(O~III)}& \colhead{$T_e$ (K)}}
\startdata
LPRST79 & Hubble~V & $ 137 $ & $11,200$ \\
PES80 & Hubble~V & $ 142\pm31 $ & $11,000\pm900$\\
STM89 & Hubble~V & $ 138\pm47 $ & $11,500\pm1,000$\\
M96   & Hubble~V & $ 110\pm26 $ & $12,400\pm1,100$\\
\hline
VT83 & I~Zw~36   & $ 70\pm12$ & $14,600\pm1,300$\\
ITL97 & I~Zw~36 & $ 59.08\pm0.47 $ & $16,180\pm65$\\
\enddata
\end{deluxetable}

\clearpage

\begin{deluxetable}{lcccc}
\tablewidth{0pc}
\tablecaption{Ionic and Total Abundances\tablenotemark{a}}
\tablehead{
\colhead{} & \multicolumn{2}{c}{Hubble~V} & \multicolumn{2}{c}{I~Zw~36}\\
\colhead{Abundance Ratio} &  \colhead{Optical} &  \colhead{SWS} &
\colhead{Optical} & \colhead{SWS}}
\startdata
O$^{+}$/H & 3,800$\pm$900 & \nodata & $710\pm80$ & \nodata\\
O$^{+2}$/H & 13,000$\pm$3,000 & \nodata & $4,930\pm40$ & \nodata\\
O/H & 16,800$\pm$3,100 & \nodata & $5,640\pm90$ & \nodata\\
\hline
S$^{+}$/H & $ 20.1\pm4.8$ & \nodata & $11.4\pm0.6$ & \nodata\\
S$^{+2}$/H & $ 250\pm90 $ &  $280\pm50$ & $ 95\pm9$ & $151\pm30$\\
S$^{+3}$/H & \nodata & $65\pm11$ & \nodata & $71\pm13$\\
\hline
S/H\tablenotemark{b} & \nodata & $365\pm51$ & \nodata & $233\pm33$\\
\hline
$\eta$ & 3.6$\pm$2.0 & \nodata & 1.2$\pm$0.2 & \nodata \\
\enddata
\tablenotetext{a}{All values are $\times10^{-8}$}
\tablenotetext{b}{S/H $=$ N(S$^+$) (from optical data) $+$
N(S$^{+2}$ $+$ S$^{+3}$) (from ISO data)}
\end{deluxetable}

\clearpage

\begin{deluxetable}{lcccc}
\tablewidth{0pc}
\tablecaption{Emission Coefficient Fit Parameters\tablenotemark{a}}
\tablehead{
\colhead{Species} & \colhead{$\lambda$ (\AA)} & \colhead{$a_{\lambda}$} &
\colhead{$b_{\lambda}$} & \colhead{$c_{\lambda}$}}
\startdata
H I & H$\beta$ & $6.072\times10^{-34}$ & $-2.327\times10^{-29}$
        & $2.989\times10^{-25}$\\
H I & Br$\alpha$ & $6.318\times10^{-35}$ &
        $-2.427\times10^{-30}$ & $2.811\times10^{-26}$\\
H I & Br$\gamma$ & $2.058\times10^{-35}$ &
        $-7.889\times10^{-31}$ & $9.365\times10^{-27}$\\
\hline
$$[O~III] & $4363$ & $2.609\times10^{-30}$ &
        $-3.828\times10^{-26}$ & $1.446\times10^{-22}$\\
$$[O~III] & $4959$ & $-4.864\times10^{-31}$ &
        $3.250\times10^{-25}$ & $-1.958\times10^{-21}$\\
$$[O~III] & $5007$ & $-1.443\times10^{-30}$ &
        $9.382\times10^{-25}$ & $-5.651\times10^{-21}$\\
$$[O~III] & $51.8\mu m$ & $1.485\times10^{-30}$ &
        $-5.523\times10^{-26}$ & $1.335\times10^{-21}$\\
$$[O~III] & $87.6\mu m$ & $1.841\times10^{-30}$ &
        $-7.676\times10^{-26}$ & $1.931\times10^{-21}$\\
\hline
$$[S~II] & $4068$ & $7.652\times10^{-31}$ & $9.599\times10^{-25}$
        & $-6.249\times10^{-21}$ \\
$$[S~II] & $4076$ & $2.136\times10^{-31}$ & $3.269\times10^{-25}$
        & $-2.122\times10^{-21}$\\
$$[S~II] & $6717$ & $-2.001\times10^{-28}$ & $8.917\times10^{-24}$
        & $-3.734\times10^{-20}$\\
$$[S~II] & $6731$ & $-1.406\times10^{-28}$ & $6.454\times10^{-24}$
        & $-2.729\times10^{-20}$\\
\hline
$$[S~III] & $6312$ & $3.884\times10^{-30}$ & $5.208\times10^{-26}$
        & $-5.035\times10^{-22}$\\
$$[S~III] & $9069$ & $-2.622\times10^{-29}$ &
        $1.079\times10^{-24}$ & $-3.318\times10^{-21}$\\
$$[S~III] & $9532$ & $-1.448\times10^{-28}$ &
        $5.960\times10^{-24}$ & $-1.833\times10^{-20}$\\
$$[S~III] & $18.7\mu m$ & $3.929\times10^{-30}$ &
        $-2.613\times10^{-25}$ & $1.681\times10^{-20}$\\
$$[S~III] & $33.5\mu m$ & $6.407\times10^{-29}$ &
        $-2.293\times10^{-24}$ & $4.178\times10^{-20}$\\
\hline
$$[S~IV] & $10.5\mu m$ & $6.125\times10^{-29}$ &
        $-3.409\times10^{-24}$ & $8.859\times10^{-20}$\\
\enddata
\tablenotetext{a}{applicable for $n_e \le 100\ cm^{-3}$}
\end{deluxetable}

\clearpage

\begin{figure}
\vspace{18cm}
\includegraphics{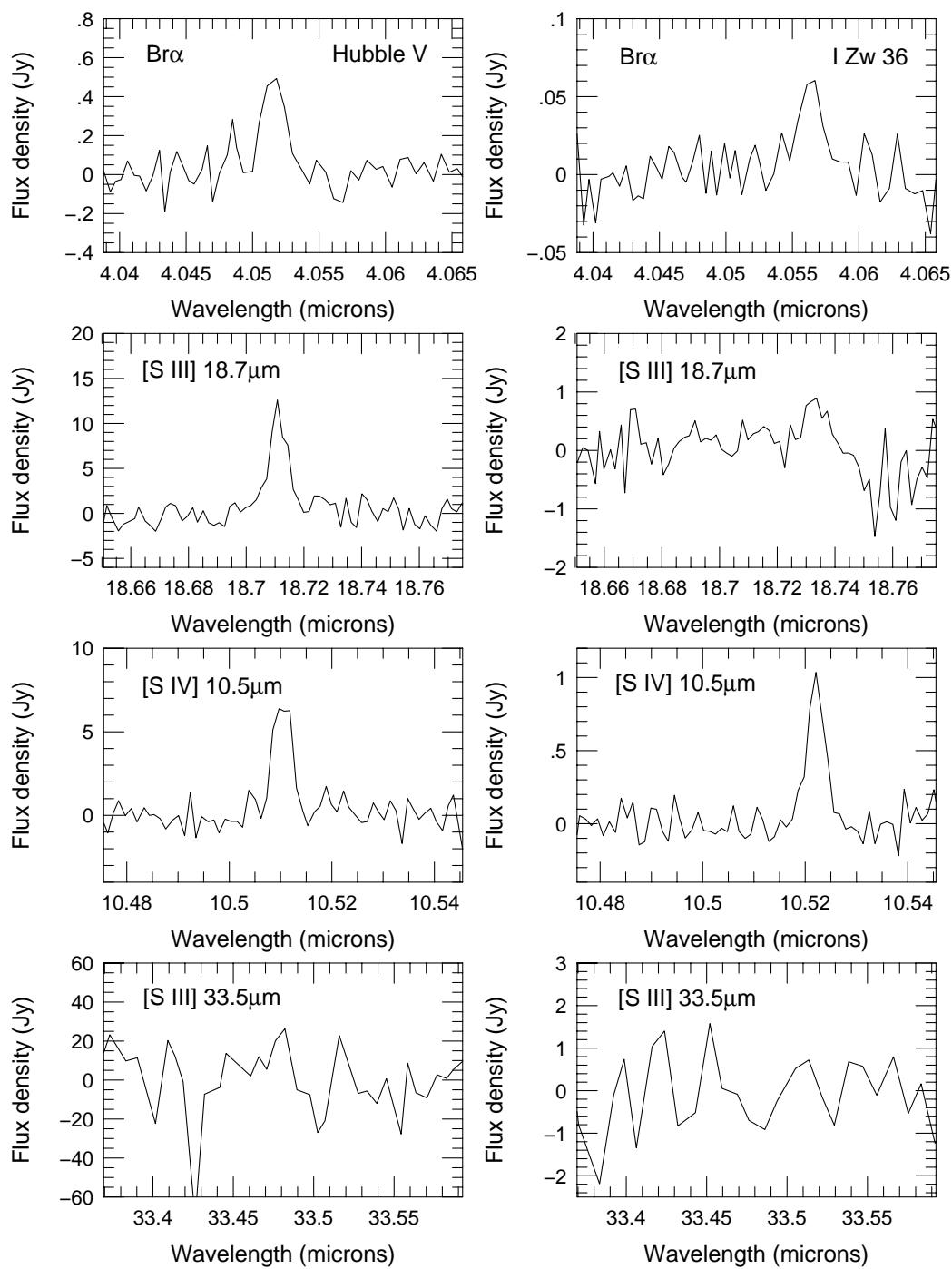}
\figcaption[Figure 1.]{$ISO$ Spectra of Br$\alpha$, [S~IV] $10.5\mu m$, 
and [S~III] $18.7$ and $33.5\mu m$ spectral lines for Hubble~V and I~Zw~36.}
\end{figure}

\clearpage

\begin{figure}
\vspace{18cm}
\includegraphics{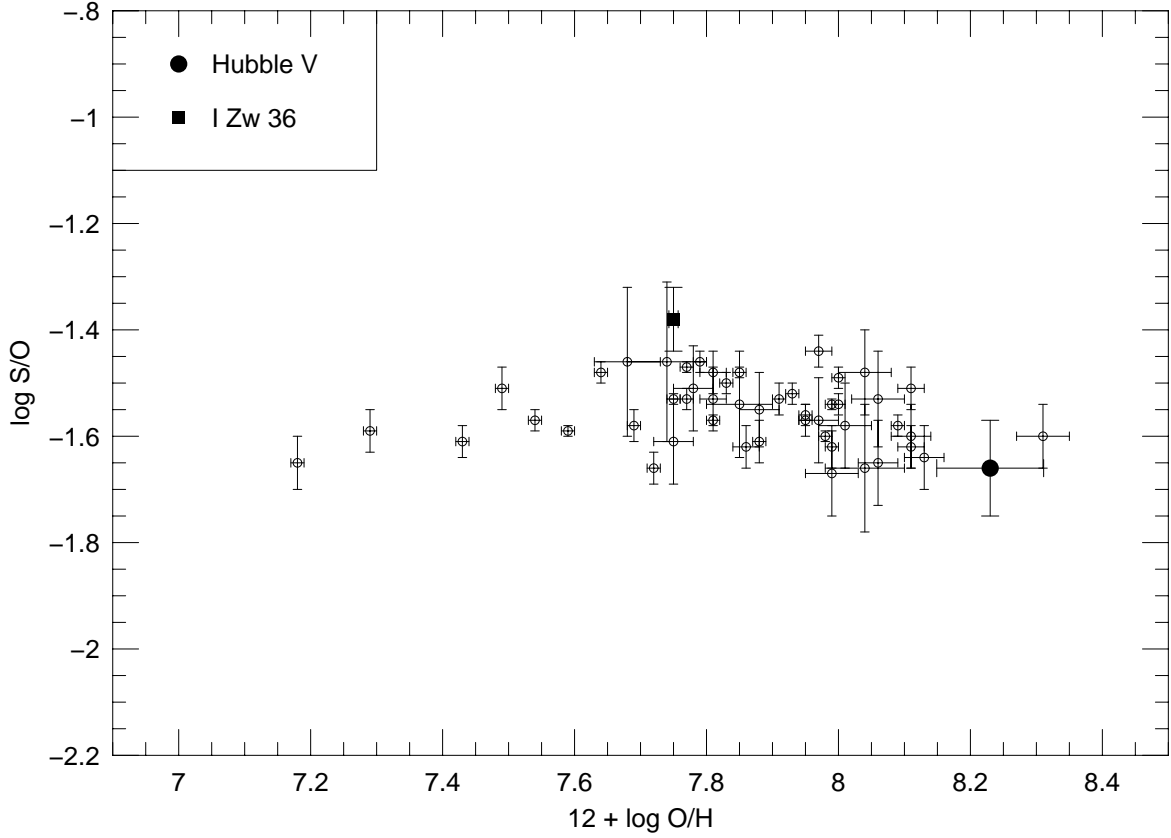}
\figcaption[Figure 2.]{The logarithm of the ratio of S/O 
plotted against the logarithm of the ratio of O/H for Hubble~V and I~Zw~36
compared to other metal-poor HII regions from Izotov \& Thuan (1999).
The S/O values for Hubble~V and I~Zw~36 include measurements of 
the S$^{+3}$ abundance from our IR observations, while the other data points
rely on ionization correction factors calculated from photoionization
modeling.}
\end{figure}

\clearpage

\begin{figure}
\vspace{18cm}
\includegraphics{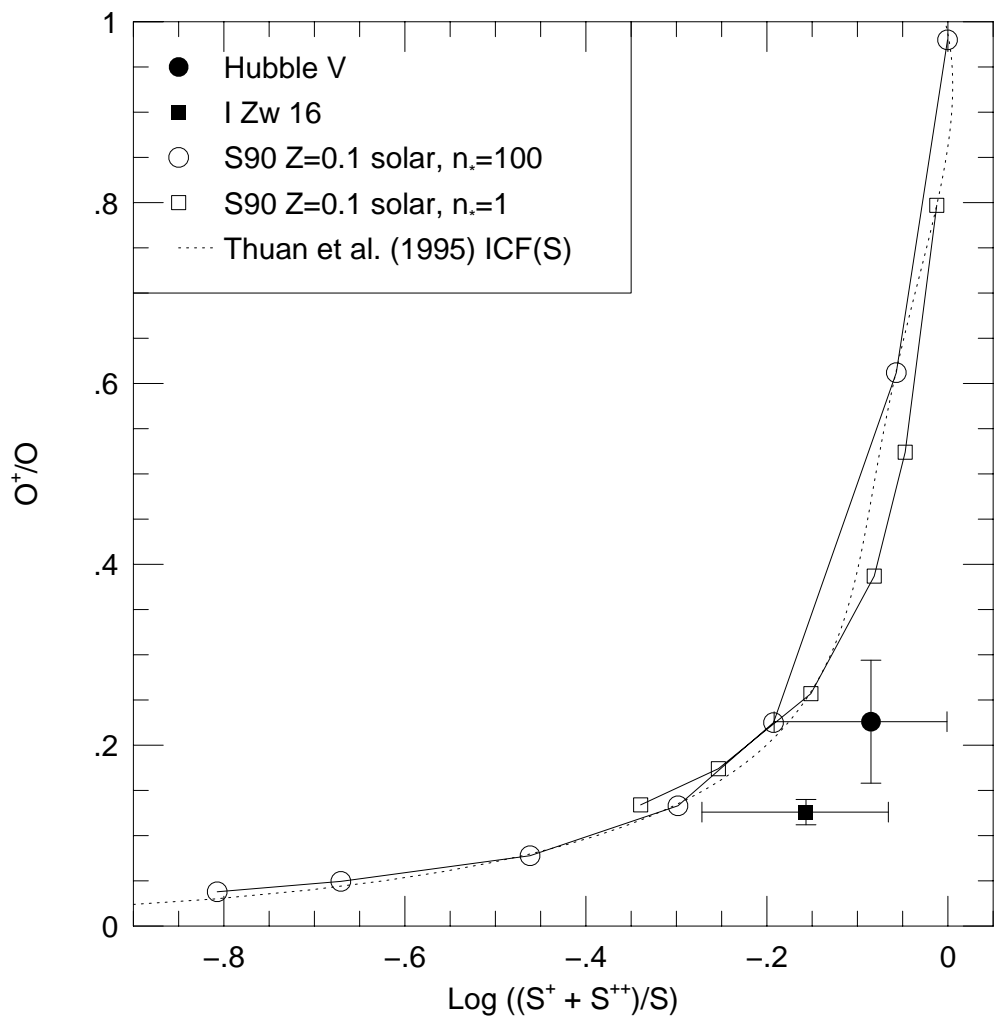}
{\small
\figcaption[Figure 3.]{ O$^{+}$/O vs. Log ((S$^{+}$+S$^{+2}$)/S) for Hubble~V 
and I~Zw~36 compared to model nebulae using different types of model 
stellar atmospheres. The filled circle and square with error bars represent 
Hubble~V and I~Zw~36, respectively. Two model sequences from Stasi\'nska (1990) 
having abundances of 0.1 solar and a different number of exciting stars 
(resulting in different ionization parameters).  The stellar effective 
temperatures range from 32,500 K to 55,000 K. 
  The fit to the models of Stasi\'nska (1990) used by 
Thuan, Izotov \& Lipovesky (1995) for the sulfur ionization correction 
factor is given by the dotted line.}}
\end{figure}

\clearpage

\begin{figure}
\vspace{18cm}
\epsscale{0.70}
\includegraphics{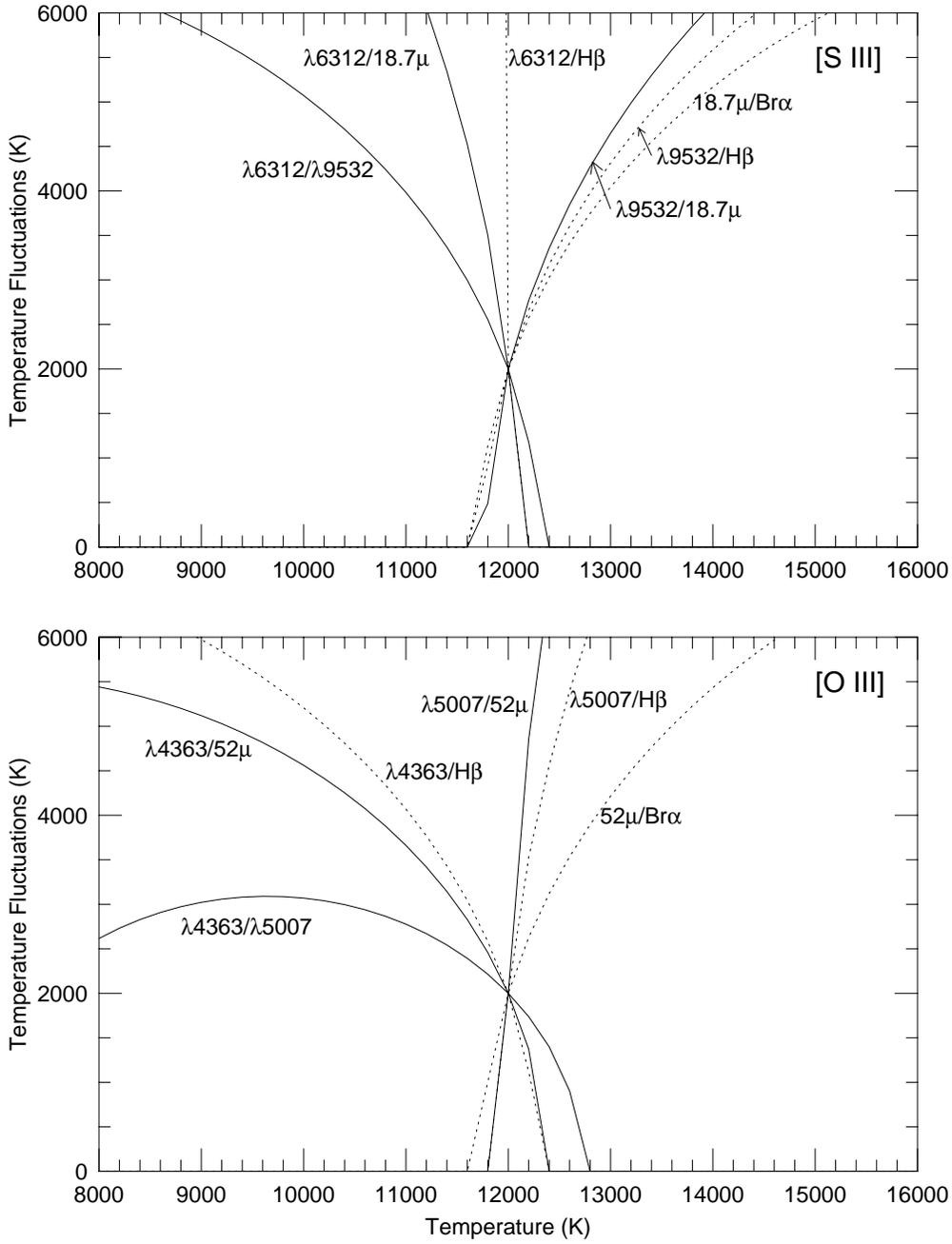} 
{\small
\figcaption[Figure 4.]{Temperature Fluctuations, $\sigma_{T}$,
for a Model Nebula with input values of $T = 12,000K$ and
$\sigma_{T} = 2,000K$.  {\it Top:}  Modelled diagnostic curves using
$N(S^{+2})/N(H) = 2.92\cdot10^{-6}$.  
{\it Bottom:}  The corresponding
graph showing diagnostic curves derived from [O~III] lines.}}
\end{figure}

\clearpage

\begin{figure}
\vspace{18cm}
\epsscale{0.70}
\includegraphics{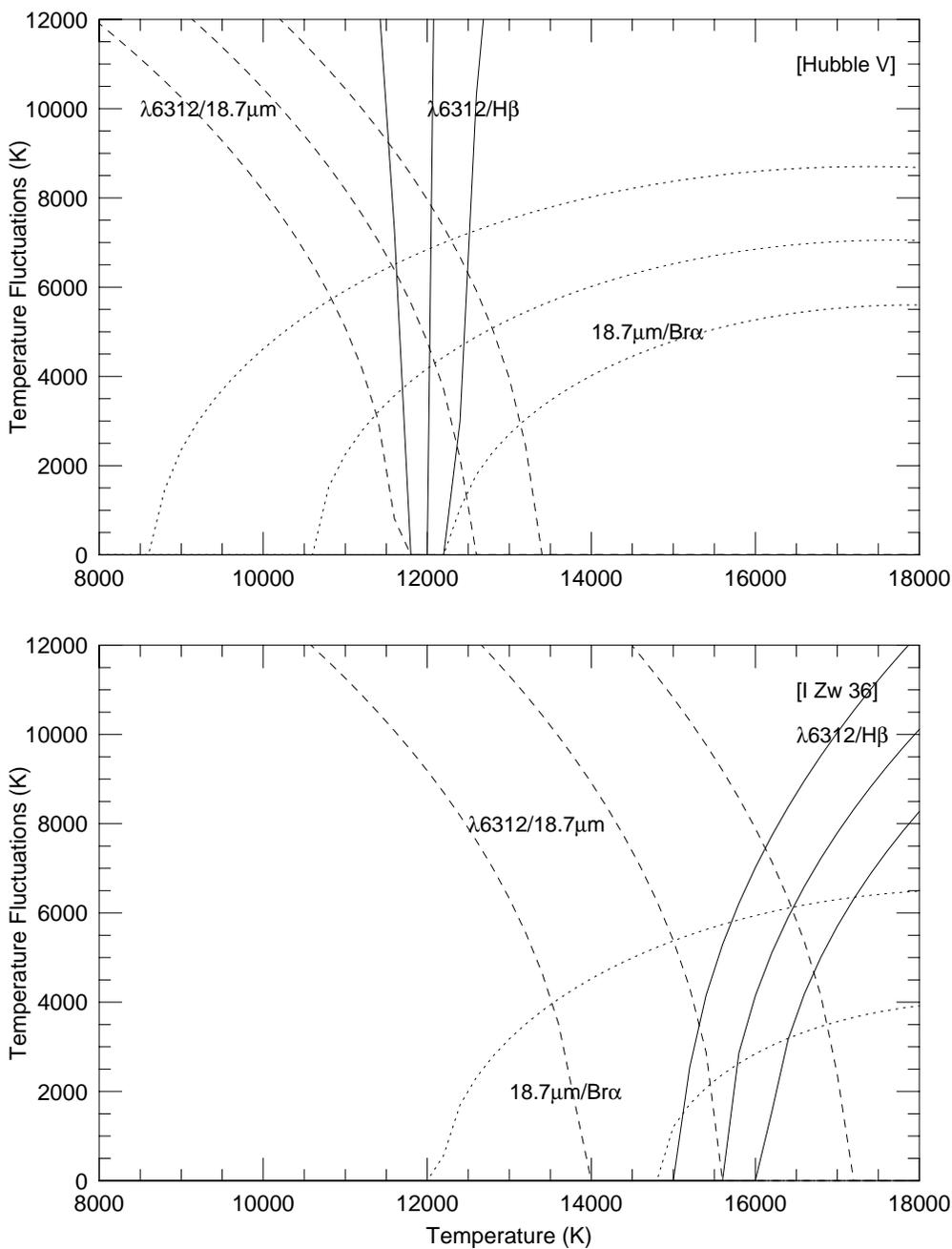} 
{\small
\figcaption[Figure 5.] {{\it Top:} A sample (see section 4.4.1)
diagnostic for Hubble~V using
line strengths from this paper as well as
Lequeux (1979). Groups of three roughly parallel lines are apparent.  The 
center line of each group is the diagnostic given by the
observed line ratios.  The other two lines correspond to the diagnostics 
given by the observed line ratios $\pm 1\sigma$. The dashed, solid, 
and dotted
lines correspond to the diagnostic functions
 from $\lambda 6312$/$18.7\mu m$, $\lambda 6312$/H$\beta$, 
and $18.7\mu m$/Br$\alpha$, respectively.
{\it Bottom:} A similar diagnostic plot for I~Zw~36 using Viallefond 
\& Thuan (1983).  }}
\end{figure}

\begin{figure}
\vspace{17cm}
\epsscale{0.5}
\includegraphics{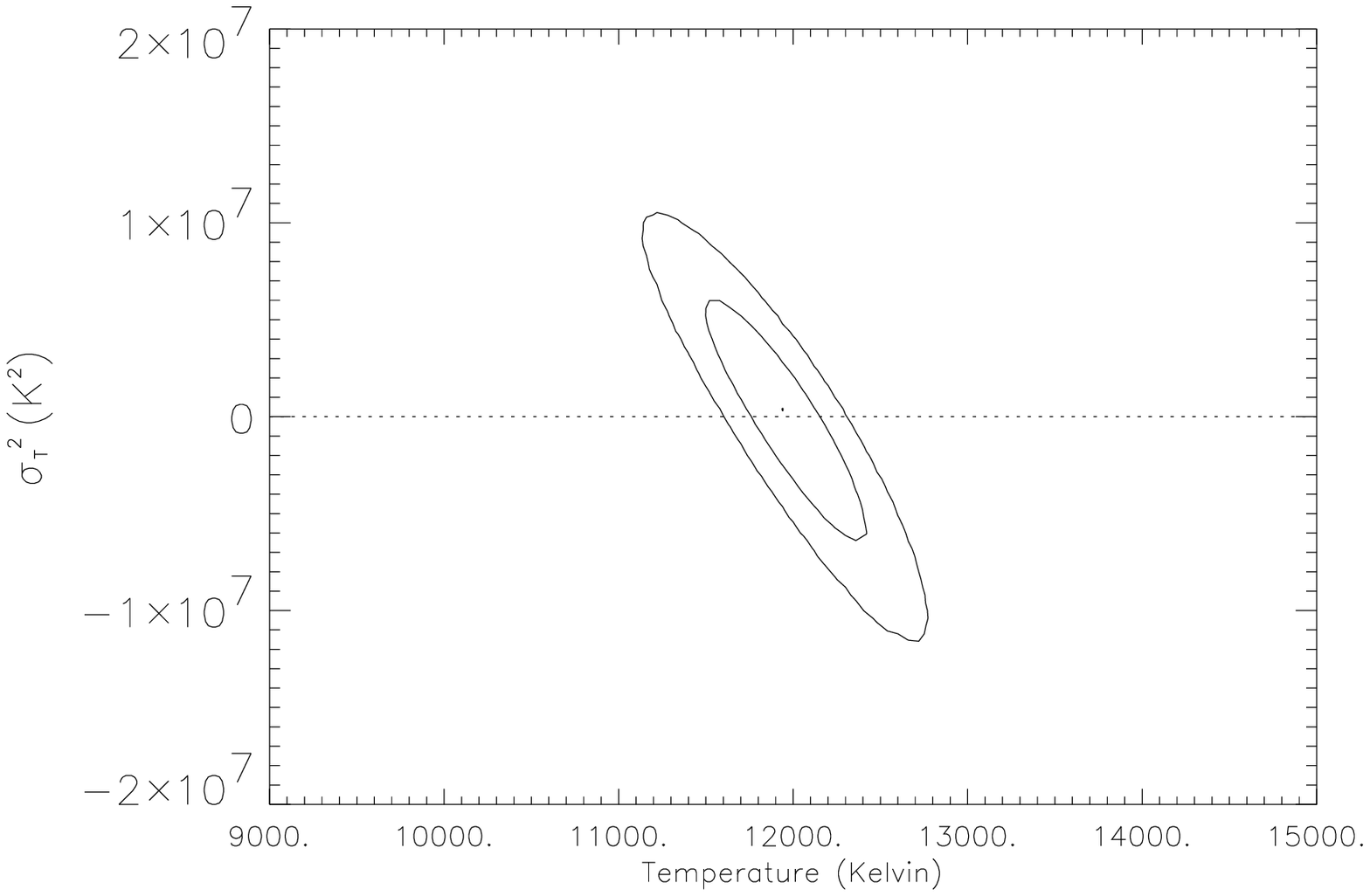} 
\includegraphics{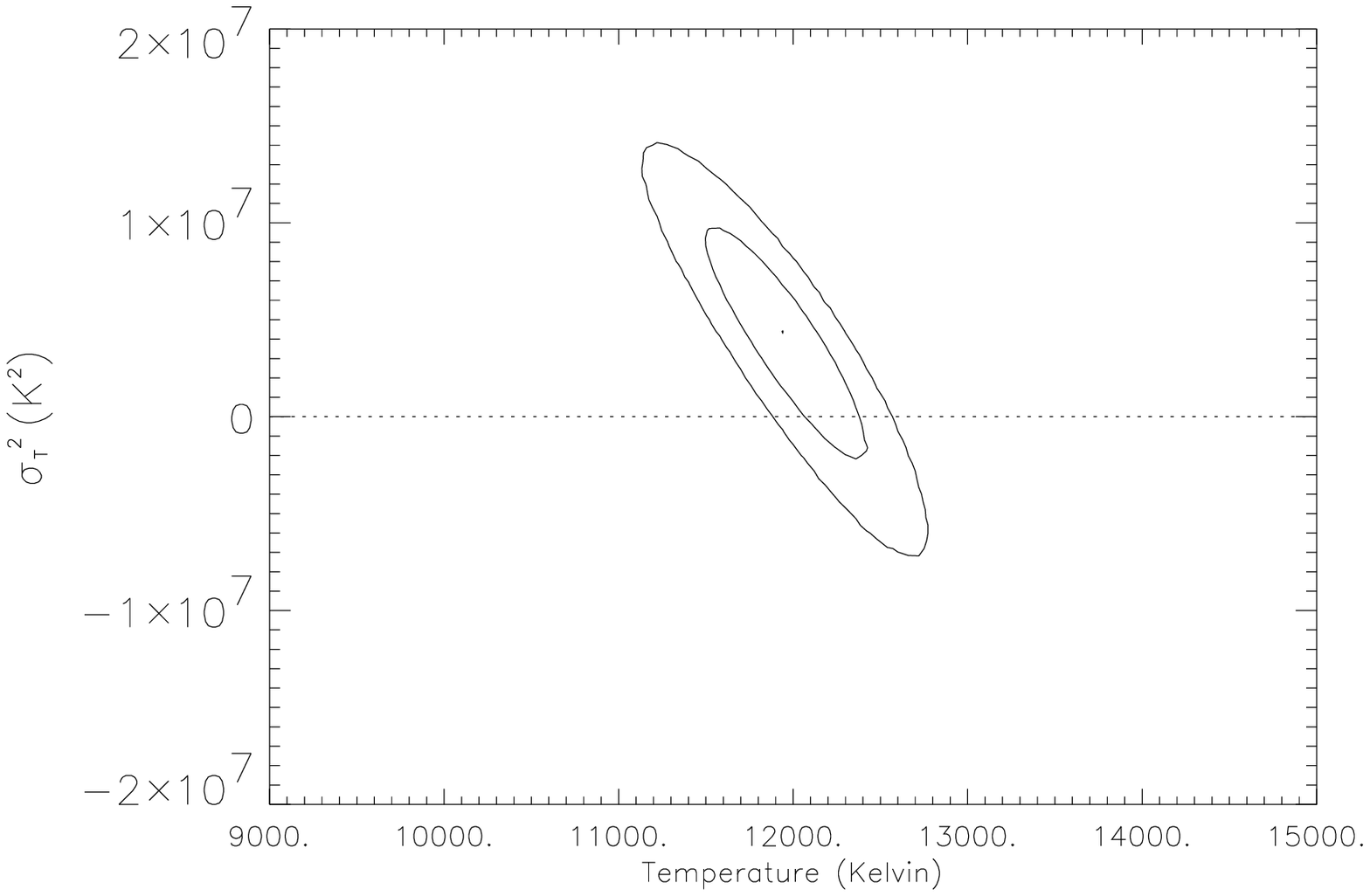} 
{\small
\figcaption[Figure 6.]{Monte Carlo Simulations showing 
values of $\sigma_{T}^{2}$ vs. $T$ derived from $2\cdot10^6$ 
simulated observations, smeared by observational errors, 
based on two different cases. Contours represent $1$, and 
$2\sigma$ confidence 
levels, while the peaks of the distributions can be seen as dots in
the center of the contours.  In each case observational errors of
$2\%$ in the line ratios [S~III] $\lambda 6312$/$\lambda 9532$ and
5\% in  $\lambda 9532$/$18.7\mu m$ were assumed.
{\it Top:} 
Simulation with a specified
nebular temperature of $T = 12,000K$ with no temperature fluctuations.
{\it Bottom:} Simulation with specified values of 
$T = 12,000K$ and $\sigma_{T} = 2,000K$, corresponding 
to $\sigma_{T}^{2} = 4\cdot10^{6}K^{2}$.}}
\end{figure}

\end{document}